\documentclass[a4]{article}
\usepackage{authblk}
\usepackage{graphicx}
\usepackage{amsmath}
\usepackage{amsfonts}
\usepackage{amssymb}

\begin{document}

%\begin{frontmatter}

\title{Van't Hoff law for temperature dependent Langmuir constants in clathrate hydrate nanocavities.}

\author{Azzedine Lakhlifi\footnote{azzedine.lakhlifi@obs-besancon.fr} \\
Institut UTINAM-UMR 6213 CNRS - Universit\'{e} de Franche-Comt\'{e} \\
Observatoire de Besan{\c c}on \\ 41 bis avenue de l'Observatoire - BP 1615 -
25010 Besan{\c c}on Cedex, France \\

Pierre R. Dahoo \\
Universit\'{e} de Versailles-Saint-Quentin-en-Yvelines, Sorbonne Universit\'{e}s \\
Laboratoire Atmosph\`{e}res Milieux Observations Spatiales, CNRS, UMR 8190 \\
Observatoire de Versailles Saint-Quentin-en-Yvelines, 
11 Bd d'Alembert, F-78820, Guyancourt, France}

\maketitle

\begin{abstract}

This work gives a van't Hoff law expression of Langmuir constants of different species for determining 
their occupancy in the nanocavities of clathrate hydrates. 
The van't Hoff law's parameters are derived  from a fit with  Langmuir constants calculated using a
pairwise site-site interaction potential to  model the  anisotropic potential environment in the cavities, as a 
function of temperature. The parameters  can be used for calculating clathrates compositions.
Results are given for nineteen gas species  trapped in the small and large cavities of 
structure types I and II \cite{lakhlifi2015}.  
The accuracy of this approach is based on a comparison 
with available experimental data for  ethane and cyclo-propane clathrate hydrates. The
numerical method applied in this work, was recently validated from a comparison with the spherical cell 
method based on analytical considerations \cite{lakhlifi2015}
  
\end{abstract}

%\begin{linenumbers}

\section{Introduction}

A clathrate is an ice-like crystalline solid consisting of water molecules forming  a cage structure 
around smaller guest molecules under suitable conditions of low temperature and high pressure. 
On Earth, it is considered that clathrate hydrates are the most important reservoirs of fossil energy 
\cite{sloan1998,sloankoh2008}, and that favourable conditions for gas hydrate formation 
exist in about 25\% of the earth's land mass. Moreover, the thermodynamics conditions of pressure 
and temperature prevailing in the oceans are such that hydrates should easily be formed in about 90\% 
of the ocean or sediments. 
The most common guest molecules in terrestrial clathrates are of organic aliphatic nature like 
methane, ethane, propane or butane, but other small inorganic molecules like 
nitrogen, carbon dioxide, and hydrogen sulfide can also  be trapped in the cages of clathrates \cite{sparkstester1992,caotestersparkstrout2001,klaudasandler2002,andersontestertrout2004,
anderson2005a,anderson2005b,sunduan2005}. 
In the advent of global warming, these clathrates can enhance the temperature rise 
when the trapped species are released. 
Clathrate hydrates are also suspected to be extensively present on several planets, satellites and comets of the 
Solar System. Planetologists are thus concerned with the possible clathrate impact on the distribution of the planet's volatiles 
and on the modification of their atmosphere's compositions \cite{sssissi2013}. Hence, it is of great interest to correctly 
determine the amount of species potentially trapped in the cages of clathrates, $i.e.$ the fractional occupancy of guest 
species under the thermodynamic conditions (pressure and temperature) prevailing  
in the regions where clathrates might form. 

From a theoretical point of view, the thermodynamics of the formation or dissociation of clathrates 
is most often based on the model developed by van der Waals and Platteeuw \cite{vdwp1959} following 
the same hypotheses under which was developed the adsorption theory of 
Langmuir \cite{langmuir1916}. The Langmuir isotherms of adsorbed molecules on a surface are 
determined from the calculation of the Langmuir constant, which is also the main parameter to be considered in 
the determination of the amount of species trapped in the clathrate cages 
as a function of pressure and temperature.

To calculate these Langmuir constants, most of the models are based on a molecular description 
of the guest-water interactions using a Lennard-Jones or Kihara potential form. 
The parameters of these potentials are usually empirically obtained from 
experimental data of phase equilibrium. Such models most often neglect interactions of the guest molecules with water 
beyond a few cages only and are therefore 
questionable \cite{thesecthomas2009,cthomas2007,cthomas2008,omousis2009,cthomas2009,tdswindle2009}. 

Moreover, it is generally assumed that the environment of 
the cage in which a gas molecule is trapped in clathrates is of spherical symmetry. Whereas this 
assumption may be justified for molecules such as CH$_4$ or NH$_3$, it is certainly not well-suited for
molecules such as CO$_2$ and N$_2$O or SO$_2$ which are of cylindrical 
or oblate symmetry and for which the hypothesis of a free rotation at the center of a spherical cage is no longer valid.

In the case of ethane and cyclo-propane clathrate hydrates, an analytical method based on the spherical 
cell model has been used to extract spherically averaged intermolecular potentials from experimental data 
on the temperature dependence of the Langmuir constant by Bazant and Trout 
\cite{bazanttrout2001}. 
For other guests in clathrate hydrates, in particular, argon, hydrogen, nitrogen, methane, ethane, 
propane, cyclo-propane, and carbon dioxide, other workers \cite{sparkstester1992,caotestersparkstrout2001,klaudasandler2002,andersontestertrout2004,sunduan2005} 
have explicitly taken into account the angle-dependence of the guest-water intermolecular potential in 
an atom-atom or site-site description to calculate the corresponding Langmuir constants. However, 
in most of these studies the water and guest molecules were simply described as 
one, two or three interaction sites for the Lennard-Jones or Kihara potential contributions although 
more than three sites are involved.

This work aims at providing a van't Hoff law expression of the Langmuir constant for 
single guest molecules incorporated in clathrate hydrates as a function of the temperature by improving 
the potential model used in the determination of the Langmuir constants 
, that is by using an atom-atom and site-site potential and by considering explicitly the 
effect of water molecules beyond the trapping cage and the resulting anisotropic environment. 
In the present work, the Langmuir constant is determined by taking into account all the external degrees 
of freedom of the guest molecules, $i.e.$ the center of mass (c.m.) translational motion and the orientational motion 
in a true crystallographic clathrate lattice, not necessarily of spherical symmetry as it is often assumed when using 
the van der Waals and Platteeuw model \cite{vdwp1959}.
In the following, we recall in section 2 the model used for the calculations of the 
Langmuir constants that are necessary to determine the fractional 
occupancy of guest species in clathrates. Then, in section 3, the geometry and interaction potential 
considered here are described. Finally, in section 4,
the coefficients for a simple van't Hoff expression of the Langmuir constants are given
for a large set of guest molecules. The Langmuir constants calculated using this model are 
compared with available experimental data, $i.e.$,  for ethane, cyclo-propane guest molecules. 

\section{Statistical thermodynamic approach}

In contrast to natural ice which solidifies in the hexagonal structure, clathrate hydrates form, as water 
crystallizes, in the cubic system in several different structures which are characterized by
specific cages of different sizes. The two most common types are "structure I" and "structure II". 
In structure I, the unit cell is made of 46 water molecules forming 
2 small (12 pentagonal faces 5$^{12}$) and 
6 large (12 pentagonal and 2 hexagonal faces 5$^{12}$6$^2$) cages, 
while in structure II, the unit cell is made up of 136 water 
molecules forming 16 small (12 pentagonal faces 5$^{12}$) and 
8 large (12 pentagonal and 4 hexagonal faces 5$^{12}$6$^4$) cages 
\cite{sloankoh2008}. 

Calculations of the relative abundances of guest species incorporated in a clathrate lattice 
structure of type I (sI) or II (sII) at given temperature-pressure conditions can be performed  
using classical statistical mechanics which allows the macroscopic thermodynamic properties 
of the clathrates to be determined from the interaction energies between the guest species 
and the clathrate water molecules. 

In 1959 van der Waals and Platteeuw \cite{vdwp1959} developed 
a model of clathrate's formation in which the trapping of guest molecules 
in nano-cages was considered to be a generalized case of the three-dimensional ideal localized adsorption. 

Their model is based on the following hypotheses:

\begin{enumerate}
\item The contribution of the host molecules to the free energy is independent of the occupational 
mode in the cages. This implies in particular that the guest species do not distort the trapping cage.

\item The encaged molecules are localized in the cages, each of which can never hold more 
than one guest. 

\item The mutual interaction of the guest molecules is neglected, i.e., the partition function for the motion 
of a guest molecule in its cage is independent of the other guests.

\item Classical statistics is valid, i.e., quantum effects are negligible.

\end{enumerate}

From the configuration partition function and the thermodynamic equilibrium condition on the chemical potentials 
of the guest and host molecules in coexisting phases in clathrate \cite{luninestevenson1985}, the 
fractional occupancy of a guest molecule $K$ in a given  
"structure type-cage size" $t$ ($t$~=~structure-type I or II-small or large cage) can be written as:

\begin{equation}
y_{K,t}=\frac{C_{K,t}f_K}{1+\sum_{J}C_{J,t}f_J},
\label{occup}
\end{equation}

\noindent where the sum in the denominator includes all the species present in 
the initial gas phase, $C_{K,t}$ is the Langmuir constant of species $K$ in the structure type-cage size $t$, 
and $f_K$ is the fugacity of the species $K$ which depends on the total pressure $P$ of the initial gas 
phase and on the temperature $T$. 

The Langmuir constant in equation (1) depends on the temperature $T$ and on the 
strength of the interaction energy between the guest species $K$ and the water molecules in the cage. 
It is expressed as:

\begin{equation}
C_{K,t}=\frac{1}{k_B
T}\int\exp\Big(-\frac{V_{K,t}(\textbf{r},\mathbf{\Omega})}{k_B T}\Big)d\textbf{r}d\mathbf{\Omega}.
\label{clang}
\end{equation}

In this equation $V_{K,t}(\textbf{r},\mathbf{\Omega})$ is the interaction potential energy 
experienced by the guest molecule for a given position vector $\textbf{r}$ of its center of 
mass with respect to the cage center and its 
orientational vector $\mathbf{\Omega}$, and $k_B$ is the Boltzmann constant. 
The integral value must be calculated for all external degrees of freedom 
of the guest molecule inside the structure type-cage size $t$.

To evaluate the Langmuir constant $C_{K,t}$, two additional assumptions are often made, namely: ($i$) the symmetry of the guest molecule's 
environment is considered to be spherical and ($ii$) the guest molecule can freely rotate in the corresponding spherical cage (spherical cell potential approximation), 
in accordance with the Lennard-Jones and Devonshire \cite{ljd1937} theory applied to liquids. 

Then, the Langmuir constant can be cast as: 

\begin{equation}
C_{K,t}=\frac{4\pi}{k_B
T}\int_{0}^{R_c}\exp\Big(-\frac{V_{K,t}(r)}{k_B T}\Big)r^2dr,
\label{clangapprox}
\end{equation}

\noindent where $R_c$ is the radius of the spherical cage and 
$V_{K,t}(r)$ is the spherically averaged potential energy between the guest molecule and 
the clathrate water molecules.

Note that Eq. (3) is commonly used by planetologists and as a result, it may introduce 
significant inaccuracies in the evaluation of the relative abundances of gas species 
in clathrate hydrates when the assumed spherical symmetry of the interaction 
potential is questionable.

In contrast, the Langmuir constant for enclathrated single guest molecules 
is determined here by taking into account all the external degrees of freedom of the enclathrated 
molecules, i.e. the center of mass (c.m.) translational motion and the orientational motion 
in a real crystallographic clathrate lattice, not necessarily of spherical symmetry. As a consequence, 
Eq. (3) is no longer valid in our approach and should be replaced by 
a more  general form, as explained below.

\section{Potential energy model} 

Let us consider a non-vibrating single gas species (atom or molecule) trapped in a small or large cage 
of a clathrate structure of type sI or sII. The 
positions of the hydrogen and oxygen atoms are given by their coordinates in a reference 
frame fixed to the clathrate lattice, while those of the guest molecule are usually given in a frame tied 
to the molecule with the \textbf{z} axis coinciding with the rotational axis of highest symmetry as shown 
for the trapping of a guest molecule (CH$_4$ for instance) in the small cage of the clathrate 
structure type I, in Figure 1.  

\begin{figure}[!ht]
\begin{center}
\includegraphics[width=12cm]{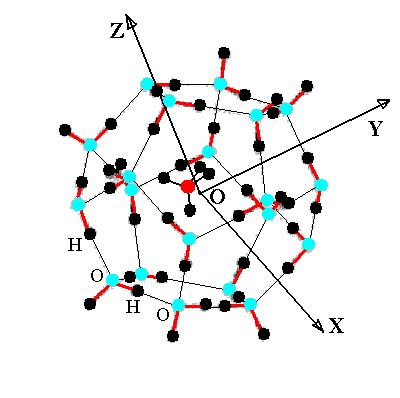}
\end{center}
\caption{Trapping geometry of a guest molecule (CH$_4$ for instance) in a 
small cage of the structure type I. 
(O,\textbf{X},\textbf{Y},\textbf{Z}) represents the absolute frame tied to the clathrate lattice.} 
\label{FIG. 1}
\end{figure}

The interaction potential energy $V_{K,t}$ between the guest molecule $K$ 
and the surrounding water molecules of the structure type-cage size $t$, considered as a 
rigid clathrate crystal, is modeled as a sum of a 12-6 Lennard-Jones (LJ) pairwise 
atom-atom potential characterizing the repulsion-dispersion contributions and an 
electrostatic part due to charge-charge interactions between the charges in the 
guest molecule and those in the water molecules of the clathrate system. 
It is expressed as: 

\begin{eqnarray}
V_{K,t} =\sum\limits_{k=1}^{N_C}\sum\limits_{j=1}^{N_W}\sum\limits_{i=1}^{N_K}
\left[4\epsilon_{ij}\left\{ \left(
\frac{\sigma _{ij}}{\left| \mathbf{r}_{ij_k}\right| }\right) ^{12} 
-\left( \frac{\sigma _{ij}}{\left| \mathbf{r}_{ij_k}\right|}\right) ^6 \right\} 
+\frac{1}{4\pi \epsilon_0} \frac{q_{i}q_{j}}
{\left| \mathbf{r}_{ij_k}\right| } \right], 
\label{Vkt}
\end{eqnarray}

\noindent where $\textbf{r}_{ij_k}$ is the distance vector between the \textit{i}th site, 
with electric charge $q_i$, 
of the guest molecule ($N_K$ sites) and the \textit{j}th site, with electric charge $q_j$, of 
the \textit{k}th water molecule ($N_W$ sites) of the clathrate matrix (containing $N_C$ water molecules); 
$\epsilon_{ij}$ and $\sigma _{ij}$ are the mixed LJ potential 
parameters, using the Lorentz-Berthelot combination rules as: 

\begin{equation}
\epsilon_{ij}=\sqrt{\epsilon_{i}\epsilon_{j}} \ \ \ \ \ \ \ \text{and}\ \ \ \ \ \ \ \sigma_{ij}=\frac{\sigma_{i}+
\sigma_{j}}{2}, 
\label{lorberth}
\end{equation}

\noindent where $\epsilon_{i}$ and $\sigma_{i}$ are the Lennard-Jones parameters 
for the $i-i$ interacting atomic pair, which are taken from the literature \cite{lakhlifigirardet1984, 
walespopelierstone1995,ketkokamathpotoff2011}. The effective electric charges for the water molecules 
$q_H$ = + 0.4238 $e$, $q_O$ = - 0.8476 $e$ are from reference \cite{alaviripmeester2010} 
and those for the guest species are from references \cite{alaviripmeester2010,ketkokamathpotoff2011,
spackman1986,pidkokazanskii2005,muenter1991}.

\begin{table}[!ht] 
\begin{center}
\caption{Pure Lennard-Jones potential parameters$^{(a)}$ used in our calculations.} 
\begin{tabular}{||c|c|c|c|c|c|c|c|c|c||} 
\hline \hline
 & & & & & & & & & \\ 
 & H - H & O - O & C - C & N - N & S - S & Ne - Ne & Ar - Ar & Kr - Kr & Xe - Xe \\
 & & & & & & & & & \\
\hline
 & & & & & & & & & \\
$\epsilon_{i}$ (K)  & 8.59 & 57.41 & 42.88 & 37.29 & 73.79 & 43.16 & 126.90 & 179.85 & 226.03 \\
 & & & & & & & & & \\
$\sigma_{i}$ (\text{\AA}) & 2.810 & 3.030 & 3.210 & 3.310 & 3.390 & 2.764 & 3.405 & 3.650 & 3.970 \\
 & & & & & & & & & \\
\hline \hline
\end{tabular}
\end{center}
\raggedright $(a)$ Obtained from Refs. \cite{lakhlifigirardet1984, 
walespopelierstone1995,ketkokamathpotoff2011}.
\label{epssig}
\end{table}

Calculations of the Langmuir constant $C_{K,t}$ from Eq. (2) require an explicit 
determination of the external degrees of freedom of the guest molecule, that is, its center of 
mass (c.m.) position vector $\textbf{r}$ and its orientation vector 
$\mathbf{\Omega}$ = ($\varphi, \theta, \chi $). 
Therefore, we define an absolute frame (O,\textbf{X},\textbf{Y},\textbf{Z}) connected to 
the clathrate matrix. 

Figure 2 gives the geometrical characteristics of the guest-water interacting 
molecules, and a description of the internal positions of the sites in the
guest molecule with respect to its frame (G,\textbf{x},\textbf{y},\textbf{z}) 
and its external orientational and translational degrees of freedom with respect 
to the absolute frame (O,\textbf{X},\textbf{Y},\textbf{Z}).

\begin{figure}[!ht]
\begin{center}
\includegraphics[width=18cm]{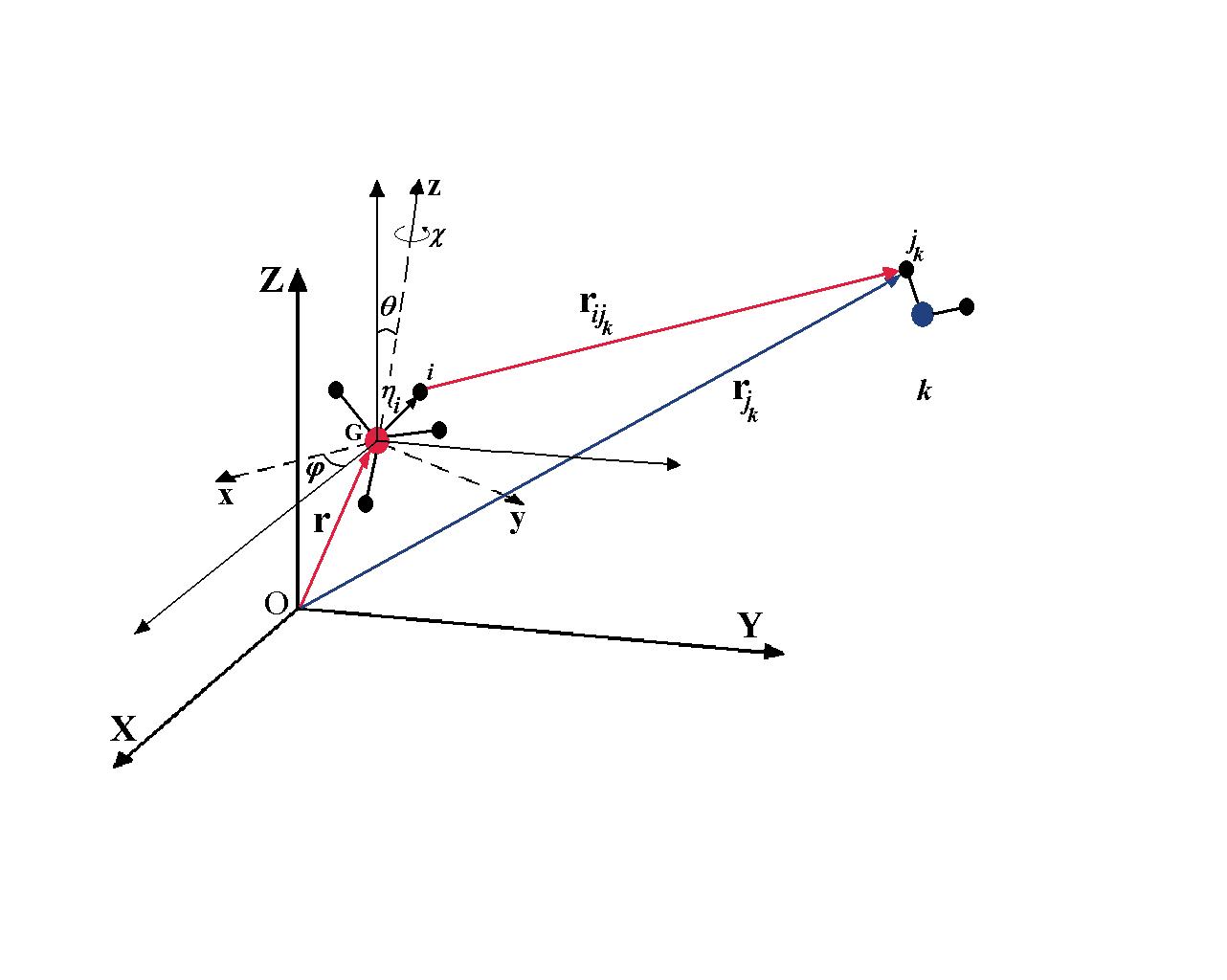}
\end{center}
\caption{Geometrical characteristics of a guest molecule (CH$_4$ for instance) interacting 
with a water molecule of the clathrate matrix. 
(O,\textbf{X},\textbf{Y},\textbf{Z}) and (G,\textbf{x},\textbf{y},\textbf{z}) represent the absolute 
frame tied to the clathrate matrix and the molecular frame, respectively.} 
\label{FIG. 2}
\end{figure}

Then, the distance vector $\mathbf{r}_{ij_k}$ in Eq. (4) can be expressed in 
terms of the position vector \textbf{r} of the c.m. of the guest molecule 
and $\mathbf{r}_{j_k}$ of the $j$th site of the $k$th water molecule with respect 
to the absolute frame (O,\textbf{X},\textbf{Y},\textbf{Z}), and of the position 
vector $\mathbf{\eta}_i$ of the $i$th site of the guest molecule with respect to its 
associated frame (G,\textbf{x},\textbf{y},\textbf{z}) (see Figure 2), as: 

\begin{equation}
\mathbf{r}_{ij_k} = \mathbf{r}_{j_k}-\mathbf{r}-\mathbf{\eta}_i.
\label{Rijk}
\end{equation}

It should be noted that the explicit dependence upon the angular degrees of freedom 
$\varphi$, $\theta$, and $\chi$ of the guest molecule requires the determination of the site 
position vectors \{$\mathbf{\eta}$\} with respect to the absolute frame.

Then, by assuming ($i$) that the guest species and water molecules 
are rigid, and ($ii$) that the clathrate lattice is undistorted (static lattice), 
the Langmuir constant is calculated from a six-dimensional configurational integral written as:

\begin{equation}
C_{K,t}=\frac{1}{k_BT}
\int\exp\Big(-\frac{V_{K,t}(x,y,z,\varphi,\theta,\chi)}{k_B T}\Big)dx dy dz d\varphi \sin\theta d\theta d\chi.
\label{clangatom}
\end{equation}

Of course, it should  be noted that ($i$) for atomic species, there are no orientational variables, and ($ii$) for 
linear molecules there is no spinning variable $\chi$.

\section{Results and discussion}

As already described above, we herein use a pairwise atom-atom Lennard-Jones and a 
site-site electrostatic potentials for calculating the interaction of single 
guest molecules with clathrate matrices containing 4 $\times$ 4 $\times$ 4 unit cells 
(up to 2944 water molecules) for the structure sI, and 3 $\times$ 3 $\times$ 3 unit cells 
(up to 3672 water molecules) for the structure sII. Indeed, it has been previously shown that these 
number of water molecules are large enough to ensure a good convergence of the corresponding 
calculations \cite{Thomas2010}. 
Moreover, in order to determine the Langmuir constant and its temperature dependence 
for a guest molecule $K$ in a structure type-cage size $t$ of clathrate, the potential energy 
surface $V_{K,t}(x,y,z,\varphi,\theta,\chi)$ in Eq. (7) is first numerically computed for 
all the center of mass positions and orientations of the molecule with respect to the 
absolute frame as defined, and then the numerical integration is performed.
Note that the minimum of the potential energy $V_{K,t}^{\text{min}}$ and the associated molecular 
position and orientation can also be determined in this way. 

\subsection{Calculation and fit of the Langmuir constants}

A clathrate is characterized by its degree of occupancy $i.e.$ the fraction of cage sites 
that is occupied by a guest molecule, which can be calculated from the knowledge of the Langmuir constant $C_{K,t}$. 
Because such constants are not necessarily extracted from experiments in a large range of temperatures,
we propose here to express them as a function of temperature through a simple van't Hoff law : 

\begin{equation}
C_{K,t}(T)=A_{K,t}\exp(B_{K,t}/T),
\label{clangvanhoff}
\end{equation}
 
\noindent in which $A$ (Pa$^{-1}$) and $B$ (K) are constant values fitted to values derived from an atomic 
description of the clathrate systems.
This simple van't Hoff expression could thus be readily used, for instance, by planetologists in the large 
range of temperatures encountered in the Solar System. 

In this work Langmuir constants have been calculated for eighteen species, in the temperature range between 
50 K and 300 K, as given in Tables 2, 3 and 4 for rare gas atoms, linear molecules and non-linear 
molecules, respectively. 

\begin{table}[!ht] 
\begin{center}
\caption{Parameters $A_{K,t}$ (Pa$^{-1}$) and $B_{K,t}$ (K) 
for the van't Hoff expression of the Langmuir constant 
for simple guest clathrate hydrates.} 
\begin{tabular}{||l|c|c|c|c||}
\hline \hline
 & & & & \\
Structure type & sI & sI & sII & sII \\
Cage size & small cage & large cage & small cage & large cage \\
 &  & & &  \\ 
\hline
 & & & & \\
Guest species $K$ & $A_{K,t}$ & $A_{K,t}$ & $A_{K,t}$ & $A_{K,t}$  \\
 & $B_{K,t}$ & $B_{K,t}$ & $B_{K,t}$ & $B_{K,t}$ \\
 & & & & \\
\hline 
 & & & & \\
Ne & 1.0308$\times 10^{-9}$  & 2.7194$\times 10^{-9}$ & 0.8154$\times 10^{-9}$ & 5.6151$\times 10^{-9}$ \\ 
       & 1187.948 & 1015.862 & 1233.898 & 898.062 \\
 &  &  &  & \\
Ar & 1.5210$\times 10^{-10}$  & 7.7829$\times 10^{-10}$ & 1.0456$\times 10^{-10}$ & 2.4531$\times 10^{-10}$  \\ 
       & 2961.545 & 2521.758 & 2977.025 & 2195.964 \\
 &  &  &  &  \\
Kr & 0.5985$\times 10^{-10}$  & 3.8799$\times 10^{-10}$ & 0.4004$\times 10^{-10}$ & 16.2508$\times 10^{-10}$  \\  
     & 3885.383 & 3454.028 & 3789.957  & 3021.690  \\
 &  &  &  &  \\
Xe & 1.9389$\times 10^{-11}$  & 13.8509$\times 10^{-11}$ & 1.2913$\times 10^{-11}$ & 83.2357$\times 10^{-11}$  \\  
      & 4547.654 & 4572.732 & 4085.506 & 4103.672 \\
 &  &  &  &  \\ 
\hline \hline
\end{tabular}
\end{center}
\label{Atoms}
\end{table}

\begin{table}[!ht] 
\begin{center}
\caption{Parameters $A_{K,t}$ (Pa$^{-1}$) and $B_{K,t}$ (K) 
for the van't Hoff expression of the Langmuir constant 
for simple guest clathrate hydrates.} 
\begin{tabular}{||l|c|c|c|c||}
\hline \hline
 & & & & \\
Structure type & sI & sI & sII & sII \\
Cage size & small cage & large cage & small cage & large cage \\
 &  & & &  \\ 
\hline
 & & & & \\
Guest species $K$ & $A_{K,t}$ & $A_{K,t}$ & $A_{K,t}$ & $A_{K,t}$  \\
 & $B_{K,t}$ & $B_{K,t}$ & $B_{K,t}$ & $B_{K,t}$ \\
 & & & & \\
\hline 
 & & & & \\
H$_2$ & 4.7301$\times 10^{-9}$ & 16.0695$\times 10^{-9}$ & 5.5295$\times 10^{-9}$ & 64.0074$\times 10^{-9}$ \\
      & 1265.757 & 1515.721 & 1203.620 & 873.259 \\
 &  &  &  &  \\
O$_2$ & 1.3153$\times 10^{-9}$ & 7.4606$\times 10^{-9}$ & 0.8004$\times 10^{-9}$ & 25.8421$\times 10^{-9}$ \\
      & 2917.693 & 2558.746 & 3044.536 & 2238.052 \\
      &  &  &  &  \\
N$_2$ & 3.9496$\times 10^{-10}$ & 25.6897$\times 10^{-10}$ & 4.8836$\times 10^{-10}$ & 201.3238$\times 10^{-10}$ \\
      & 2869.400 & 2680.372 & 2679.423 & 2226.480 \\
      &  &  &  &  \\
CO & 2.5937$\times 10^{-10}$ & 16.5331$\times 10^{-10}$ & 4.5198$\times 10^{-10}$ & 229.8631$\times 10^{-10}$ \\
      & 3518.021 & 3075.059 & 3088.930 & 2275.803 \\
      &  &  &  &  \\
CO$_2$ & 7.7765$\times 10^{-12}$ & 520.5579$\times 10^{-12}$ & 7.9970$\times 10^{-12}$ & 6907.0012$\times 10^{-12}$ \\
     & 2976.629 & 4674.690 & 2277.757 & 3370.363 \\
     &  &  &  &  \\
HCN & 7.6653$\times 10^{-12}$  & 131.1607$\times 10^{-12}$  & 13.9141$\times 10^{-12}$  & 8224.5133$\times 10^{-12}$  \\
     & 4085.369 & 4328.556 & 2593.031 & 2640.868 \\
     &  &  &  &  \\
C$_2$H$_2$  & 0.9702$\times 10^{-12}$  & 221.9622$\times 10^{-12}$  & 0.2439$\times 10^{-12}$  & 3909.8421$\times 10^{-12}$  \\
     & 735.205 & 3076.356 & 321.114 & 2837.467 \\
 &  &  &  &  \\
\hline \hline
\end{tabular}
\end{center}
\label{linearmol}
\end{table}

\begin{table}[!ht] 
\begin{center}
\caption{Parameters $A_{K,t}$ (Pa$^{-1}$) and $B_{K,t}$ (K) 
for the van't Hoff expression of the Langmuir constant 
for simple guest clathrate hydrates.} 
\begin{tabular}{||l|c|c|c|c||}
\hline \hline
 &  &  &  &  \\
Structure type & sI & sI & sII & sII \\
Cage size & small cage & large cage & small cage & large cage \\
 &  & & &  \\ 
\hline
 &  &  &  &  \\
Guest species $K$  & $A_{K,t}$ & $A_{K,t}$ & $A_{K,t}$ & $A_{K,t}$  \\
 & $B_{K,t}$ & $B_{K,t}$ & $B_{K,t}$ & $B_{K,t}$ \\
  &  &  &  &  \\
\hline 
 & & & & \\
H$_2$S & 2.3444$\times 10^{-10}$ & 7.2080$\times 10^{-10}$ & 3.6415$\times 10^{-10}$ & 758.3575$\times 10^{-10}$ \\
     & 4463.910 & 4073.045 & 3073.324 & 2495.937 \\
     &  &  &  &  \\
SO$_2$ & 1.2311$\times 10^{-12}$  & 75.0641$\times 10^{-12}$  & 6.1515$\times 10^{-12}$  & 17926.7530$\times 10^{-12}$  \\
     & 4374.084 & 6272.810 & 1548.504 & 4139.948 \\
     &  &  &  &  \\
NH$_3$ & 8.6697$\times 10^{-11}$ & 27.5220$\times 10^{-11}$ & 38.7320$\times 10^{-11}$ & 5031.4000$\times 10^{-11}$ \\
     & 5197.361 & 4975.753 & 3334.132 & 2484.181 \\
     &  &  &  &  \\
CH$_4$ & 8.3453$\times 10^{-10}$ & 116.6313$\times 10^{-10}$ & 5.4792$\times 10^{-10}$ & 829.8039$\times 10^{-10}$ \\
     & 2901.747 & 2959.901 & 2546.660 & 2629.194 \\
     &  &  &  &  \\
C$_2$H$_6^{(\text a)}$  & - & 3.5164$\times 10^{-11}$ & - & 727.2717$\times 10^{-11}$ \\
    & - & 4226.997 & - & 4440.484 \\
     &  &  &  &  \\
cyc-C$_3$H$_6^{(\text a)}$ & - & 1.4881$\times 10^{-11}$ & - & 402.4295$\times 10^{-11}$ \\
    & - & 4781.938 & - & 5161.620 \\
    &  &  &  &  \\
C$_3$H$_8^{(\text a)}$ & - & 5.5707$\times 10^{-13}$ & - & 597.9850$\times 10^{-13}$ \\
    & - & 3537.025 & - & 7118.782 \\
    &  &  &  &  \\
iso-C$_4$H$_{10}^{(\text a)}$& - & 2.7970$\times 10^{-14}$ & - & 208.3210$\times 10^{-14}$ \\
    & - & 1598.004 & - & 7103.169 \\
    & &  &  &  \\
\hline \hline
\end{tabular}
\end{center}
\raggedright $(a)$ When the potential energy surfaces are positive, the Langmuir constants 
are negligible and the parameters $A_{K,t}$ and $B_{K,t}$ are not given.
\label{Nonlinearmol}
\end{table}

\subsection{Comparison with available experimental Langmuir constants}

\subsubsection{Ethane and cyclo-propane simple hydrates}         

In Tables 5 and 6, the Langmuir constants calculated here are compared to 
experimental values obtained from the dissociation pressure data by Sparks and Tester 
\cite{sparkstester1992} for temperatures ranging between 200 K and 290 K and 
between 240 K and 290 K, for ethane and cyclo-propane trapped in large cages of 
structure sI, respectively. Moreover, for comparison the Langmuir constants 
calculated from a simple spherical model as often used in the literature are also reported 
in Tables 5 and 6.

\begin{table}[!ht] 
\begin{center}
\caption{Calculated and experimental Langmuir constants (Pa$^{-1}$) for ethane 
trapped in large cage of structure sI. Experimental values are obtained from 
reference \cite{sparkstester1992}.} 
\begin{tabular}{||c|c|c|c||}
\hline \hline
 & & & \\ 
$T$ (K) & $C_{\text{cal.}}$ this work & $C_{\text{exp.}}$ & $C_{\text{cal.}}$ Spherical model \\
 & & & \\
 \hline 
  & & & \\
200 & 5.3 $\times 10^{-2}$ & 9.9 $\times 10^{-2}$ & 14.7 $\times 10^{-2}$ \\
 & & & \\
210 & 2.0 $\times 10^{-2}$ & 3.2 $\times 10^{-2}$ & 5.2 $\times 10^{-2}$ \\
 & & & \\
220 &  7.8 $\times 10^{-3}$ & 12.0 $\times 10^{-3}$ & 20.2 $\times 10^{-3}$ \\
 & & & \\
230 & 3.4 $\times 10^{-3}$ & 4.6 $\times 10^{-3}$ & 8.5 $\times 10^{-3}$ \\
 & & & \\
240 &  1.6 $\times 10^{-3}$ & 1.9 $\times 10^{-3}$ & 3.9 $\times 10^{-3}$ \\
 & & & \\
250 & 7.7 $\times 10^{-4}$ & 8.4 $\times 10^{-4}$ & 18.6 $\times 10^{-4}$ \\
 & & & \\
260 & 4.0 $\times 10^{-4}$ & 4.1 $\times 10^{-4}$ & 9.5 $\times 10^{-4}$ \\
 & & & \\
270 & 2.2 $\times 10^{-4}$ & 2.0 $\times 10^{-4}$ & 5.1 $\times 10^{-4}$ \\
 & & & \\
280 & 1.3 $\times 10^{-4}$ & 1.1 $\times 10^{-4}$ & 2.9 $\times 10^{-4}$ \\
 & & & \\
290 & 7.5 $\times 10^{-5}$ & 6.1 $\times 10^{-5}$ & 16.7 $\times 10^{-5}$ \\
 & & & \\
\hline \hline
\end{tabular}
\end{center}
\label{Ethane}
\end{table}

\begin{table}[!ht] 
\begin{center}
\caption{Calculated and experimental Langmuir constants (Pa$^{-1}$) for cyclo-propane 
trapped in large cage of the sI structure. 
Experimental values are obtained from reference \cite{sparkstester1992}.} 
\begin{tabular}{||c|c|c|c||}
\hline \hline
 & & & \\ 
$T$ (K) & $ C_{\text{cal.}} $ this work & $ C_{\text{exp.}}$ & $ C_{\text{cal.}}$ Spherical model \\
 & & & \\
 \hline 
  & & & \\
240 & 6.7 $\times 10^{-3}$ & 9.5 $\times 10^{-3}$ & 0.12 $\times 10^{-3}$ \\
 & & & \\
250 & 3.0 $\times 10^{-3}$ & 3.8 $\times 10^{-3}$ & 0.06 $\times 10^{-3}$ \\
 & & & \\
260 & 1.5 $\times 10^{-3}$ & 1.7 $\times 10^{-3}$ & 0.04 $\times 10^{-3}$ \\
 & & & \\
270 & 7.3 $\times 10^{-4}$ & 7.6 $\times 10^{-4}$ & 0.23 $\times 10^{-4}$ \\
 & & & \\
280 & 3.9 $\times 10^{-4}$ & 3.8 $\times 10^{-4}$ & 0.15 $\times 10^{-4}$ \\
 & & & \\
290 & 2.2 $\times 10^{-4}$ & 2.0 $\times 10^{-4}$ & 0.10 $\times 10^{-4}$ \\
 & & & \\
\hline \hline
\end{tabular}
\end{center}
\label{cycpropane}
\end{table}

It can be seen that the values calculated here are in good agreement with those extracted from experimental 
data. The ratios $C_{\text{cal}} / C_{\text{exp}}$ vary between 0.5 and 1.2 for ethane, 
and between 0.7 and 1.1 for cyclo-propane, depending on temperature. Note that the same ratios calculated 
within the same temperature range for the large cages of structure sII lead to values ranging from 
320 to 540 for ethane, and from 920 to 1100 for cyclo-propane. These results are consistent 
with the fact that ethane and cyclo-propane form sI-large cage clathrate hydrates only.
Moreover, Tables 5 and 6 show that the best agreement
between the values calculated here and those extracted from experiments is found when temperature is  
higher than 260 K. 
Below this temperature, the calculated values are systematically lower than the experimental ones.

To extract a quantitative value for our comparison, we use the following least square deviation 
parameter $\sigma$ defined as: 

\begin{equation}
\sigma=\sqrt{\frac{\sum\limits_{i=1}^{N}(C_\text{cal}-C_{\text Y})^2}{N-1}},
\label{devparam}
\end{equation}

\noindent where $C_{\text{cal}}$ are the values calculated at different temperatures in this work 
and $C_{\text{Y}}$ are the values obtained experimentally \cite{sparkstester1992} 
or from the spherical model as often used in the literature.  
$N$ is the number of temperatures considered 
($N$ = 10 in Table 5). We find values of $\sigma$ equal to 1.6 $\times 10^{-2}$ Pa$^{-1}$ and 
3.3 $\times 10^{-2}$ Pa$^{-1}$, respectively. 
Moreover, note that in our calculations, the minimum of the potential energy corresponds to 
ethane's center of mass located at 0.12 {\AA} from the center of the large cage of structure sI, i.e., a value 
that is consistent with observations from single crystal X-ray diffraction studies that place 
the center of mass at 0.17 {\AA} from the center of the large cage \cite{udachin2002}.

As for cyclo-propane (values given in Table 6 ($N$ = 6)), we calculate values of $\sigma$ equal 
to 1.3 $\times 10^{-3}$ Pa$^{-1}$ and 3.3 $\times 10^{-3}$ Pa$^{-1}$, respectively. 

To summarize, when compared to experimental results,  the values of the Langmuir constants calculated 
here for ethane show a better agreement with experimental data than those calculated from the spherical 
model, although they are of the same order of magnitude. 

Nevertheless, for cyclo-propane Table 6 shows that values obtained from a simple spherical model 
disagree with experimental data and then they are certainly not correct. 

Therefore,  the comparison of values calculated in this work with values issued from experimental work 
provides firm and sufficient grounds to validate the results obtained here for ethane and cyclo-propane.

\section{Conclusions}

In the present work, calculations of the Langmuir constants and their temperature dependence 
for simple-guest clathrate hydrates have been performed for nineteen gas species using the 
van der Waals and Platteeuw model and an all-atom approach for calculating the interactions 
between guest and water molecules in the clathrate cages. This approach accounts for the 
atomic character of the guest species and for the non-spherical water environment. 

Then, the temperature dependence of the Langmuir constants in the range 50 K - 300 K is 
given in the form of a van't Hoff expression with parameters obtained from a fit to
calculated values.

This simple expression could thus be easily applied in various situations, especially when 
experimental data are not available.  For example, it can
help planetologists in the determination of  
the fractional occupancies of gas species trapped in the clathrates that are suspected to
exist in various solid bodies of the Solar System.
Indeed, this information is mandatory to analyze the atmospheric compositions of planets 
and satellites like Mars and Titan, for example, and thus to 
better understand their way of formation. 
Unfortunately, the temperatures usually considered in clathrate experiments are often 
far from the temperature range of interest for Planetology. As a consequence, calculations of 
clathrate occupancies is a preliminary step for any formation scenario that would consider 
clathrate influence. 
In that purpose, using the van't Hoff expression given here would be much more simple 
than the usual approaches. 

\newpage

\textbf{Acknowledgements}

This work has been carried out thanks to the support of the INSU EPOV interdisciplinary program.
A. Lakhlifi thanks Kevin van Keulen and S\'{e}kou Diakit\'{e} for fruitful discussions.

Simulations have been executed on computers from the Utinam Institute of the Universit\'{e}  de 
Franche-Comt\'{e}, supported by the R\'{e}gion de Franche-Comt\'{e} and Institut des Sciences 
de l'Univers (INSU).

\end{document}